\documentclass[12pt]{article}
\usepackage{amsmath}
\usepackage{amssymb}
\usepackage{amsfonts}
\newcommand{\ba}{\begin{array}{l}}
\newcommand{\ea}{\end{array}}
\newcommand{\beq}{\begin{equation}}
\newcommand{\eeq}{\end{equation}}
\newcommand{\bea}{\begin{eqnarray}}
\newcommand{\eea}{\end{eqnarray}}
\usepackage{epsfig}
\usepackage{graphicx}
\usepackage{color}

\definecolor{dyellow}{rgb}{1.,0.8,.0}
\definecolor{myblue}{rgb}{.1,.1,.7}
\definecolor{dcyan}{rgb}{.0,.6,.6}
\definecolor{dmagenta}{rgb}{0.6,0.0,0.6}
\definecolor{brown}{rgb}{0.6,0.2,0.}
\definecolor{darkblue}{rgb}{.0,.0,0.5}
\definecolor{darkred}{rgb}{0.75,0.0,0.0}
\definecolor{orange}{rgb}{1.,.6,.0}
\definecolor{dorange}{rgb}{0.8,.4,.0}
\definecolor{darkgreen}{rgb}{0.0,0.6,0.0}
\definecolor{purple}{rgb}{.4,.0,.4}

\def\bc{\begin{center}}
\def\ec{\end{center}}
\def\be{\begin{eqnarray}}
\def\ee{\end{eqnarray}}

\newcommand{\omits}[1]{}

\begin{document}
\begin{center}
{\Large \bf { Ultra-high energy cosmic rays threshold in Randers-Finsler space }}\\
  \vspace*{1cm}
Zhe Chang \footnote{changz@mail.ihep.ac.cn} and  Xin
Li \footnote{lixin@mail.ihep.ac.cn}\\
\vspace*{0.2cm} {\sl Institute of High Energy Physics\\
Chinese Academy of Sciences\\
P. O. Box 918(4), 100049 Beijing, China}\\

\bigskip

\end{center}
\vspace*{2.5cm}

%
\begin{abstract}\baselineskip=30pt
Kinematics in Finsler space is used to study the propagation of
ultra high energy cosmic rays particles through the cosmic microwave
background radiation. We find that the GZK threshold is lifted
dramatically in Randers-Finsler space. A tiny deformation of
spacetime from Minkowskian to Finslerian allows more ultra-high
energy cosmic rays particles arrive at the earth. It is suggested
that the lower bound of particle mass is related with the negative
second invariant speed in Randers-Finsler space.
 \vspace{1cm}
\begin{flushleft}
PACS numbers:  03.30.+p, 11.30.Cp, 98.70.Sa
\end{flushleft}
\end{abstract}

\newpage
\baselineskip=30pt

Decades ago, Greisen, Zatsepin and Kuz'min (GZK) \cite{GZK}
discussed the propagation of 
the ultra-high energy cosmic rays (UHECR) particles through the
cosmic microwave background radiation (CMBR) \cite{Wilkinson}. Due
to  photopion production process by the CMBR, the UHECR particles
will lose their energies drastically down to a theoretical threshold
(about $5\times 10^{19}$eV). That is to say, the UHECR particles
which their energy beyond the threshold can not be
observed\cite{Stecker}. This strong suppression is called GZK
cutoff. However hundreds of events with energies above $10^{19}$eV
and about 20 events above $10^{20}$eV have been observed\cite{GZK
beyond}.

To explain this puzzle, one general accepted hypothesis is that the
Lorentz Invariance (LI) is violated\cite{LI}. The violation of the
LI and the Planck scale physics have long been suggested as possible
solutions of the cosmic rays threshold anomalies\cite{LI}. LI is one
of the foundations of the Standard model of particle physics.
Coleman and Glashow have set up a perturbative framework for
investigating possible departures of local quantum field theory from
LI\cite{Coleman1,Coleman2}. In a different approach, Cohen and
Glashow suggested \cite{Glashow} that the exact symmetry group of
nature may be isomorphic to a subgroup SIM$(2)$ of the Poincare
group. The mere observation of ultra-high energy cosmic rays and
analysis of neutrino data give an upper bound of $10^{-25}$ on the
Lorentz violation\cite{neutrino}.

In fact, Gibbons, Gomis and Pope\cite{Gibbons} showed that the
Finslerian line element $ds=(\eta_{\mu\nu} dx^\mu
dx^\nu)^{(1-b)/2}(n_\rho dx^\rho)^b$ is invariant under the
transformations of the group DISIM$_b(2)$. The very special
relativity is a Finsler geometry.

Recently, we proposed a gravitational field equation in
Berwald-Finsler space\cite{Berwald}. The asymmetric term in field
equation violated LI naturally. A modified Newton's gravity is
obtained as the weak field approximation of the Einstein's equation
in Berwald-Finsler space\cite{lixin}. The flat rotation curves of
spiral galaxies can be deduced naturally without invoking dark
matter in the framework of Finsler geometry.

In this Letter, we use the kinematics in Randers-Finsler space to
study the propagation of the UHECR particles through CMBR. We obtain
a deformed GZK threshold for the UHECR particles interacting with
soft photons, which depends on an intrinsic parameter of the
Randers-Finsler space\cite{chang}.

Denote by $T_xM$ the tangent space at $x\in M$, and by $TM$ the
tangent bundle of $M$. Each element of $TM$ has the form $(x, y)$,
where $x\in M$ and $y\in T_xM$. The natural projection $\pi :
TM\rightarrow M$ is given by $\pi(x, y)\equiv x$. A Finsler
structure\cite{Finsler} of $M$ is a function\be F :
TM\rightarrow[0,\infty)\nonumber. \ee The Finsler structure $F$ is
regularity (F is $C^\infty$ on the entire slit tangent bundle
$TM\backslash0$), positive homogeneity ($F(x, \lambda y)=\lambda
F(x, y)$, for all $\lambda>0$) and strong convexity (the $n\times n$
Hessian matrix $g_{ij}\equiv\frac{\partial^2}{\partial y^i\partial
y^j }\left(\frac{1}{2}F^2\right)$ is positive-definite at every
point of $TM\backslash0$).

It is convenient to take $y\equiv \frac{dx}{d\tau}$ being the
intrinsic speed on Finsler space.

In 1941, G.~Randers\cite{Randers} studied a very interesting class
of Finsler manifolds.  The Randers metric is a Finsler structure $F$
on $TM$ with the form \be\label{Randers metric} F(x,y)\equiv
\sqrt{\eta_{ij}\frac{dx^i}{d\tau}\frac{dx^j}{d\tau}}+\frac{\eta_{ij}\kappa^i}{2m}\frac{dx^j}{d\tau}
~.\ee

The action of a free moving particle on Randers space is given as
\be I=\int^r_s\mathcal{L}d\tau=m\int^r_s
F\left(\frac{dx}{d\tau}\right)d\tau .\ee Define the canonical
momentum $p_i$ as \be p_i=m\frac{\partial F}{\partial
\left(\frac{dx^i}{d\tau}\right)}~.\ee Using Euler's theorm on
homogeneous functions, we can write the mass--shell condition as \be
{\cal M}(p)=g^{ij}p_ip_j =m^2~.\ee

The modified dispersion relation in Randers spaces is of the form
\be\label{MDR} m^2=\eta^{ij}p_ip_j-\eta^{ij}\kappa_i(\mu,M_p)
p_j~,\ee where we have used the notation \be \eta_{ij}={\rm
diag}\{1,-1,-1,-1\}~,\\
\kappa_i=\kappa \{1,-1,-1,-1\}~, \ee and $\eta^{ij}$ is the inverse
matrix of $\eta_{ij}$. Here $\kappa$ can be regarded as a
measurement of LI violation. We consider the head-on collision
between a soft photon of energy $\epsilon$, momentum q and a high
energy particle $m_1$ of energy $E_1$, momentum $p_1$, which leads
to the production of two particles $m_2$, $m_3$ with energies $E_2$,
$E_3$ and momentums $p_2$, $p_3$, respectively. By making use of the
energy and momentum conservation law and the modified dispersion
relation (\ref{MDR}), we obtain the deformed GZK threshold in
Randers-Finsler space \be
E_{th}=\frac{(m_2+m_3)^2-m_1^2}{4(\epsilon-\kappa/2)} .\ee

Taking roughly the energy of soft photon to be $10^{-3}$eV, we give
a plot for the dependence of the threshold $E^N_{th}$ on the
deformation parameter $\kappa$ in FIG. 1.

\includegraphics[scale=1.5]{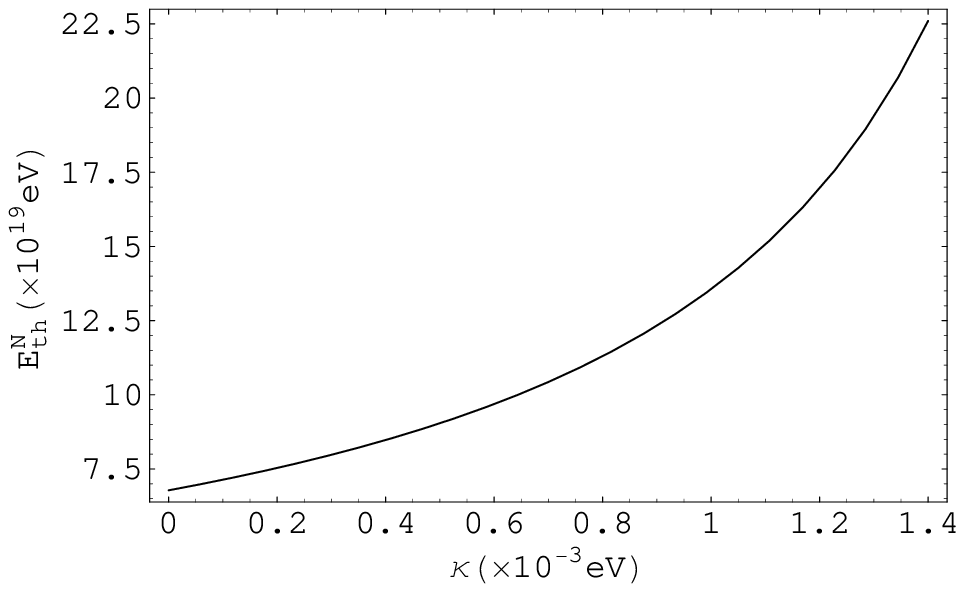}{FIG.1}

We can see clearly that a tiny deformation of spacetime ($\kappa$
with the order of the CMBR) can provide sufficient correction to the
primary predicted threshold for the propagation of UHECR particles
through the CMBR\cite{GZK}. If the nature of our universe is
Finslerian, more UHECR particles should be detected than Greisen,
Zatsepin and Kuzmin expected.

Another invariant speed in Randers-Finsler space is expressed
as\cite{chang} \be C_2=\frac{\kappa-4m}{\kappa+4m}~. \ee From the
above discussion, we know that the deformation parameter $\kappa$
may be the same order with CMBR. So far as we know that there is no
observational evidence for the existence of the second invariant
speed $C_2$. Thus, we suppose that the $C_2$ is negative or $C_2$ is
beyond the speed of light.  The negative condition of the invariant
speed $C_2$ deduces that $m\geq\kappa/4$. This gives particle mass a
lower bound for massive particle. The condition that $C_2$ is beyond
the speed of light deduces that the mass of particle is negative. In
such a case, $C_2$ may be corresponded to the speed of Goldstone
boson.

Recently, there is a renewed interest in experimental tests of LI
and CPT symmetry. Kostelecky\cite{Kostelecky} has tabulated
experimental results for LI and CPT violation in the minimal
Standard-Model Extension. Our result would not violate the minimal
Standard-Model Extension, since $\kappa$ can be eliminated by a
redefinition of the energy and momentum. $\kappa$ is very small, the
minor change in energy and momentum can be neglected except for soft
photon.

\bigskip

\centerline{\large\bf Acknowledgements} \vspace{0.5cm}
 We would like to thank T. Chen,
J. X. Lu, N. Wu, M. L. Yan and Y. Yu for useful discussion. One of
us (X. Li) indebt W. Bietenholz for useful discussion on UHECR. The
work was supported by the NSF of China under Grant NOs. 10575106 and
10875129.


\begin{thebibliography}{999}
\bibitem{GZK}K. Greisen, Phys. Rev. Lett. {\bf 16}, 748 (1966); G. T. Zatsepin and V. A. Kuzmin,
JETP Lett. {\bf 4}, 78 (1966).
\bibitem{Wilkinson}P. G. Roll and D. T. Wilkinson, Phys. Revs. Lett.
{\bf 16}, 405 (1966).
\bibitem{Stecker}F. W. Stecker, Phys. Rev. Lett. {\bf 21}, 1016 (1968).
\bibitem{GZK beyond}M. Takeda {\it et al}., Phys. Rev. Lett. {\bf 81}, 1163
(1998);M. Takeda et al., Astrophys. J. 522, 225 (1999); N. Hayashida
{\it et al}., Phys. Rev. Lett. {\bf 73}, 3491 (1994); D. J. Bird et
al., Astrophys. J. {\bf 441}, 144 (1995); D. J. Bird {\it et al}.,
Phys. Rev. Lett. {\bf 71}, 3401 (1993); D. J. Bird {\it et al}.,
Astrophys. J. {\bf 424}, 491 (1994); M. A. Lawrence, R. J. O. Reid,
and A. A. Watson, J. Phys. G {\bf 17}, 733 (1991).
\bibitem{LI}V. A. Kostelecky, Phys. Rev. D. {\bf 69}, 105009
(2004); R. Aloisio, P. Blasi, P. L. Ghia, and A. F. Grillo, Phys.
Rev. D {\bf 62}, 053010 (2000); O. Bertolami and C. S. Carvalho,
Phys. Rev. D {\bf 61}, 103002 (2000); H. Sato, arXiv:
astro-ph/0005218; T. Kifune, Astrophys. J. {\bf 518}, L21 (1999).
 W. Kluzniak, arXiv: astro-ph/9905308;  R. J. Protheroe and
H. Meyer, Phys. Lett. B {\bf 493}, 1 (2000).
\bibitem{Coleman1}S.R. Coleman and S.L. Glashow, Phys. Lett. B{\bf
                  405}, 249 (1997).
\bibitem{Coleman2}S.R. Coleman and S.L. Glashow, Phys. Rev. D{\bf
                  59}, 116008 (1999).
\bibitem{Glashow} A.G. Cohen and S.L. Glashow, Phys. Rev. Lett. {\bf
                  97} 021601 (2006).
\bibitem{neutrino}G. Battistoni {\em et al.}, Phys. Lett. B{\bf 615}
14 (2005).
\bibitem{Gibbons}G.W. Gibbons, J. Gomis and C.N. Pope, "{\em General
Very Special Relativity is Finsler Geometry}", hep-th/0707.2174.
\bibitem{Berwald}X. Li and Z. Chang, arXiv: gr-qc/0711.1934.
\bibitem{lixin}Z. Chang and X. Li, arXiv: gr-qc/0806.2184, to be
published in Phys. Lett. B.
\bibitem{chang}Z. Chang and X. Li, Phys. Lett. B. {\bf 663}, 103
(2008).

\bibitem{Finsler}D. Bao, S.S. Chern and Z. Shen, {\em An
Introduction to Riemann-Finsler Geometry}, Graduate Texts in
Mathematics {\bf 200}, Springer, New York, 2000.
\bibitem{Randers}G. Randers, Phys. Rev. {\bf 59}, 195 (1941).
\bibitem{Kostelecky}V. A. Kostelecky, arXiv: hep-ph/0801.0287v1.
\end{thebibliography}
\end{document}